\documentclass[preprint,12pt]{aastex}

\newcommand{\beq}{\begin{equation}}           
\newcommand{\eeq}{\end{equation}}             
\newcommand{\bef}{\begin{figure}}           
\newcommand{\enf}{\end{figure}}             

\shorttitle{Light Deflection by Moving Microlenses}

\begin{document}

\title{Velocity Effects on the Deflection of Light by Gravitational Microlenses}

\author{David Heyrovsk\'y}

\affil{Institute of Theoretical Physics, Charles University\\ V Hole\v{s}ovi\v{c}k\'ach 2, 18000 Prague, Czech Republic}
\email{david.heyrovsky@mff.cuni.cz}

\begin{abstract}
We study the influence of general lens and source velocities on the gravitational deflection of light by single and two-point-mass microlenses with general axis orientation. We demonstrate that in all cases the lens equation preserves its form exactly. However, its parameters -- the Einstein radius and the binary-lens separation -- are influenced by the lens velocity. In Galactic microlensing settings the velocity mainly affects the inferred separation for wide binary-star or star+planet microlenses oriented close to the line of sight. We briefly discuss the case of lenses moving with highly relativistic velocities.

\end{abstract}


\section{INTRODUCTION}

The time-dependence of Galactic gravitational microlensing events \citep{pacz96} is given by the relative motion of the source of light, the lens, and the observer near perfect alignment. These events have been generally successfully analyzed using a quasi-static approach, assuming that light is deflected at each instant in a static source-lens-observer configuration. This approach is well justified -- the lens and source velocities relative to the observer are of the order of a couple hundred $km\,s^{-1}$, i.e., $\sim 10^{-3}$ of the speed of light $c$. 

Nevertheless, with the introduction of image subtraction techniques \citep{allu98} the accuracy of measured microlensing light curves has increased in ideal cases to sub-percent levels. In addition there are prospects of high-precision observations of the astrometric microlensing effect (see \citealt{bsv98}; \citealt{han01} for the single- and binary-lens cases, respectively) by space-based interferometers such as the Space Interferometry Mission (SIM)\footnote{\url{http://sim.jpl.nasa.gov}}. Given these developments, one can expect that corrections to the light curve or the angular image geometry down to the order $\sim10^{-3}$ might be detectable. It is therefore interesting to avoid the quasi-static approach and investigate the effect of general lens and source velocities on light deflection.

Previous theoretical research has mostly concentrated on the effect of a relative lens to observer velocity on the deflection angle by a single lens. \cite{pybi93} derived the first-order effect for a velocity of arbitrary direction, showing that the deflection angle increased for lenses moving away from the observer and decreased for those moving towards the observer. The same result was later confirmed by \cite{frit03}. Recently \cite{wusp04} presented results for an arbitrarily large but purely radial lens velocity. The most general theoretical result can be found in the detailed treatise by \cite{kosch99}, which includes a derivation of the deflection angle for an ensemble of lenses of general velocities.  

In this paper we study light deflection for general lens and source velocities (arbitrarily oriented and arbitrarily large) specifically for single and two-point-mass lens microlensing events. In \S2 we derive the general lens equation and demonstrate the velocity dependence of its parameters. In \S3 we explore the magnitude of the velocity effects for low velocities and illustrate the highly relativistic limit. We conclude in \S4 by discussing observational aspects and the appropriateness of some of the assumptions.

\section{DERIVATION OF LENS EQUATION}

Our approach is similar to the one used by \cite{klio03} for a single lens in motion. We utilize the knowledge of light-deflection formulae in the rest-frame of the lens and using Lorentz transformations connect the solution to the rest-frame of the observer. We limit our accuracy to the usual first order in deflection angle. For example light rays passing close to the components of a binary pulsar are thus beyond the scope of this paper.

In the case of two-point-mass lenses (hereafter ``binary lenses" for brevity -- includes binary-star and star + planet lens systems) we concentrate on the effect of their center-of-mass velocity and neglect the effect of orbital velocity. This is justifiable for sufficiently wide binaries (with semi-major axis $\gtrsim 1 AU$). We return to the case of closer binaries in \S4. 

For the purposes of the following calculation we set up the observer rest-frame coordinates with the origin at the center of mass of the lens at time $t'=0$, the $z'$ axis pointing towards the observer, the $x'$ axis in the plane of the sky along the projected binary-lens axis, and the $y'$ axis perpendicular to it in the plane of the sky. We denote the distance between the observer and the (center of mass of the) lens $D_L$, the distance between the observer and the source plane $D_S$, and the distance between the lens and source planes $D_{LS}\equiv D_S-D_L$.

For simplicity we scale all velocities to the speed of light. We denote the lens and source velocities measured in the rest-frame of the observer \boldmath$V$ and $W$\unboldmath, respectively. At time $t'$ in the described observer rest-frame the observer, the center of mass of the lens, and the source are located at
\beq
\begin{array}{l}
\mbox{\boldmath$r$}'_O=(0,0,D_L) \\
\mbox{\boldmath$r$}'_L(t')=\mbox{\boldmath$V$}t' \\
\mbox{\boldmath$r$}'_S(t')=\mbox{\boldmath$r$}'_{S0}+ \mbox{\boldmath$W$}(t'+D_{LS})\;,
\end{array}
\label{eq:positions}
\eeq
respectively. In the last expression $\mbox{\boldmath$r$}'_{S0}=(\beta_1 D_S, \beta_2 D_S, -D_{LS})$ is the source position at $t'=-D_{LS}$. The two-dimensional angle $\mbox{\boldmath$\beta$}$ is the angular position of the source in the plane of the sky. Note that an undeflected photon arriving at the observer at $t'=D_L$ passed the lens at $t'=0$ and the source at $t'=-D_{LS}$. The lens and source velocities and distances are thus measured at these retarded times.

We denote the total mass of the lens $M$; in the binary-lens case the two lensing bodies have masses $M_A\equiv\mu_A M$ and $M_B\equiv\mu_B M$, respectively. As hinted earlier, we place no restrictions on the mass ratio of the two lenses -- our results thus hold for single- and binary-star lenses, as well as for star+planet lenses. We define the coordinates in the rest-frame lens of the lens with the origin at the center of mass of the lens, the $x$ axis along the physical binary-lens axis, the $y$ axis in the plane of the sky parallel to $y'$, and the $z$ axis perpendicular to both. In these coordinates the lenses are located at $\mbox{\boldmath$r$}_A=(-\mu_B R,0,0)$ and $\mbox{\boldmath$r$}_B=(\mu_A R,0,0)$, where $R$ is the distance between the two lenses (the intrinsic separation). We denote the angle between the binary-lens axis and the plane of the sky $\zeta$, oriented so that a small positive value brings lens A closer to the observer. The directions of the $x$ and $z$ axis thus coincide with those of the observer's $x'$ and $z'$ only if the binary-lens axis lies in the plane of the sky, i.e., $\zeta=0$. 

Denoting the future asymptotic trajectory of a photon (at the observer) in lens-frame coordinates
\beq
\mbox{\boldmath$r$}_{\!po}(t)=\mbox{\boldmath$l$}+\mbox{\boldmath$n$}_0 t\;,
\label{eq:asympobs}
\eeq
where $\mbox{\boldmath$n$}_0$ is a unit vector, the past asymptotic trajectory of the photon (at the source) is
\beq
\mbox{\boldmath$r$}_{\!ps}(t)=\mbox{\boldmath$r$}_{\!po}(t)+\frac{4G}{c^2} \sum_{i=A,B}M_i\,\frac{\mbox{\boldmath$n$}_0\mbox{\boldmath$\cdot$} [\mbox{\boldmath$r$}_{\!po}(t)-\mbox{\boldmath$r$}_i]} {[\mbox{\boldmath$n$}_0\mbox{\boldmath$\times$} (\mbox{\boldmath$l$}-\mbox{\boldmath$r$}_i)]^2} \,[\,\mbox{\boldmath$l$}-\mbox{\boldmath$r$}_i-\mbox{\boldmath$n$}_0 \mbox{\boldmath$\cdot$}(\mbox{\boldmath$l$}-\mbox{\boldmath$r$}_i)\, \mbox{\boldmath$n$}_0]\;,
\label{eq:asympsource}
\eeq
as demonstrated for example in \cite{will81} or \cite{brum91}. To obtain the values of the constant vectors $\mbox{\boldmath$l$}$ and $\mbox{\boldmath$n$}_0$, we transform the future asymptotic trajectory to observer-frame coordinates, in which the light ray arriving at the observer is
\beq
\mbox{\boldmath$r$}'_{\!po}(t')=\mbox{\boldmath$r$}'_O+\mbox{\boldmath$n$}'_0 (t'-D_L)\;.
\eeq
The unit vector $\mbox{\boldmath$n$}'_0=(1+\theta^2)^{-1/2} (-\theta_1,-\theta_2,1)$ describes the direction of the light ray at arrival. The two-dimensional angle $\mbox{\boldmath$\theta$}$ denotes the angular position of the image in the plane of the sky.

The conversion between the two coordinate systems is given by
\beq
\left( \begin{array}{c} t' \\ \mbox{\boldmath$r$}'_{\!po}(t') \end{array} \right) = \mathrm{\Lambda}(\mbox{\boldmath$V$}) \left( \begin{array}{c} t \\ \mathrm{R}(\zeta)\,\mbox{\boldmath$r$}_{\!po}(t) \end{array} \right)\;,
\label{eq:transobserver}
\eeq
where the rotation matrix
\beq
\mathrm{R}(\zeta)=\left(\! \begin{array}{ccc} \cos\zeta & 0 & \sin\zeta \\ 0 & 1 & 0 \\ -\sin\zeta & 0 & \cos\zeta \\ \end{array} \right)
\eeq
corrects for the orientation of the binary-lens axis, and the Lorentz boost matrix \cite[e.g.,][]{mtw73} is
\beq
\mathrm{\Lambda}^{0'}_{\;\;0}=\gamma,\; \mathrm{\Lambda}^{0'}_{\;\;i}=\mathrm{\Lambda}^{i'}_{\;\;0}=\gamma V_i,\; \mathrm{\Lambda}^{i'}_{\;\;j}=\delta^{ij}+ \frac{V_iV_j}{V^2}(\gamma-1)\;,
\eeq
with $\gamma=(1-V^2)^{-1/2}$. From equation~(\ref{eq:transobserver}) we express $\mbox{\boldmath$r$}_{\!po}(t)$ and by comparison with equation~(\ref{eq:asympobs}) we get the two vectors $\mbox{\boldmath$l$}$ and $\mbox{\boldmath$n$}_0$ describing the future asymptotic light ray in the lens frame. 

In a similar way we can take the expression for the past asymptotic trajectory of the photon from equation~(\ref{eq:asympsource}) and extend it to the source as follows:
\beq
\left( \begin{array}{c} t'_e \\ \mbox{\boldmath$r$}'_S(t'_e) \end{array} \right) = \mathrm{\Lambda}(\mbox{\boldmath$V$}) \left( \begin{array}{c} t_e \\ \mathrm{R}(\zeta)\,\mbox{\boldmath$r$}_{\!ps}(t_e) \end{array} \right)\;.
\label{eq:transsource}
\eeq
Note that $\mbox{\boldmath$r$}'_S(t'_e)$ depends on the source velocity $\mbox{\boldmath$W$}$ -- see equation~(\ref{eq:positions}). To get the photon position $\mbox{\boldmath$r$}_{\!ps}(t_e)$ we substitute the previously obtained light-ray vectors $\mbox{\boldmath$l$}$ and $\mbox{\boldmath$n$}_0$ into equation~(\ref{eq:asympsource}). We first use the time component of equation~(\ref{eq:transsource}) to convert between the emission times $t_e$ and $t'_e$ in the two coordinate systems, then we use the $z$ component to eliminate the emission time altogether.

From the remaining two equations we can finally express the light-deflection angle \mbox{$\mbox{\boldmath$\alpha$}\equiv (\mbox{\boldmath$\theta$}-\mbox{\boldmath$\beta$})D_S/D_{LS}$} measured by the observer. In the single lens case we get to first order in deflection angle 
\beq
\mbox{\boldmath$\alpha$}(\mbox{\boldmath$\theta$})= \frac{4GM(1-V_z)}{c^2D_L\sqrt{1-V^2}}\,\frac{\mbox{\boldmath$\theta$}} {\theta^2}\;.
\label{eq:singledefl}
\eeq
This expression is in agreement with the previous results of \citet{kosch99}, and extends the results of \citet{pybi93}, \citet{frit03}, and \citet{wusp04}. In the binary-lens case we get to first order in deflection angle
\beq
\mbox{\boldmath$\alpha$}(\mbox{\boldmath$\theta$})= \frac{4GM(1-V_z)}{c^2D_L\sqrt{1-V^2}} \left[\mu_A\frac{\mbox{\boldmath$\theta$}-\mbox{\boldmath$\theta$}_A} {|\mbox{\boldmath$\theta$}-\mbox{\boldmath$\theta$}_A|^2} +\mu_B\frac{\mbox{\boldmath$\theta$}-\mbox{\boldmath$\theta$}_B} {|\mbox{\boldmath$\theta$}-\mbox{\boldmath$\theta$}_B|^2}\right]\;,
\label{eq:binarydefl}
\eeq
where $\mbox{\boldmath$\theta$}_A\equiv-\mu_B\mbox{\boldmath$\theta$}_{AB}$ and $\mbox{\boldmath$\theta$}_B\equiv\mu_A\mbox{\boldmath$\theta$}_{AB}$ are the apparent angular positions of the two lenses, and the apparent angular separation vector (from lens A to B) in the sky is
\beq
\mbox{\boldmath$\theta$}_{AB}(\mbox{\boldmath$V$})= \frac{1}{D_L}\left\{\mbox{\boldmath$R$}_\perp+ \left[\,R_z-\frac{1\!-\!\sqrt{1-V^2}}{V^2}\,\mbox{\boldmath$V\!\cdot\!R$}\, \right] \frac{\mbox{\boldmath$V$}\!_\perp}{1-V_z}\right\}\;.
\label{eq:angularsep}
\eeq
Here the tangential lens velocity $\mbox{\boldmath$V$}\!_\perp\equiv(V_x,V_y)$ and the vector $\mbox{\boldmath$R$}\equiv(\mbox{\boldmath$R$}_\perp, R_z)=\mathsf{R}(\zeta)(\mbox{\boldmath$r$}_B-\mbox{\boldmath$r$}_A)$ is the separation vector of the two lenses in their center-of-mass rest-frame at time $t'=0$. Its component $\mbox{\boldmath$R$}_\perp$ lies in the plane of the sky and $R_z$ is oriented along the line of sight to the observer (for a positive $R_z$ lens B lies closer to the observer).

The first obvious result is the independence of the deflection angle on the source velocity $\mbox{\boldmath$W$}$. We note that we have demonstrated this only to the first order in deflection angle, the source velocity might have an influence at a higher accuracy.

The structure of equations~(\ref{eq:singledefl}) and (\ref{eq:binarydefl}) allows us to readily define the velocity-dependent angular Einstein radius
\beq
\theta_E(\mbox{\boldmath$V$})\equiv \sqrt{\frac{4GMD_{LS}}{c^2D_LD_S}\,\frac{1-V_z}{\sqrt{1-V^2}}}\;.
\label{eq:Einstein}
\eeq
We can re-scale all angular quantities by $\theta_E(\mbox{\boldmath$V$})$ and thus convert from $(\mbox{\boldmath$\beta$}, \mbox{\boldmath$\theta$}, \mbox{\boldmath$\theta$}_A, \mbox{\boldmath$\theta$}_B, \mbox{\boldmath$\theta$}_{AB})$ to $(\mbox{\boldmath$y$}, \mbox{\boldmath$x$}, \mbox{\boldmath$x$}_A, \mbox{\boldmath$x$}_B, \mbox{\boldmath$d$})$. The obtained lens equation for a single lens is
\beq
\mbox{\boldmath$y$}-\mbox{\boldmath$x$}= -\frac{\mbox{\boldmath$x$}}{x^2} \;.
\eeq
The lens equation for the binary lens is
\beq
\mbox{\boldmath$y$}-\mbox{\boldmath$x$}= -\mu_A\frac{\mbox{\boldmath$x$}-\mbox{\boldmath$x$}_A} {|\mbox{\boldmath$x$}-\mbox{\boldmath$x$}_A|^2} -\mu_B\frac{\mbox{\boldmath$x$}-\mbox{\boldmath$x$}_B} {|\mbox{\boldmath$x$}-\mbox{\boldmath$x$}_B|^2}\;.
\eeq

This interesting result shows that in either case the lens equations have exactly the same form as in the static case. When analyzing microlensing observations one can thus use exactly the same formulae as in the usual quasi-static approach. However, when interpreting the fitted parameters one has to realize that the Einstein radius, and in the binary-lens case also the lens positions and their separation, depend on the lens velocity. Instead of $\theta_E(0)$ one obtains $\theta_E(\mbox{\boldmath$V$})$ and instead of the angular lens separation vector $\mbox{\boldmath$d$}(0)\equiv \mbox{\boldmath$R$}_\perp/[D_L\theta_E(0)]$ one obtains
\beq
\mbox{\boldmath$d$}(\mbox{\boldmath$V$})= \frac{\sqrt[4]{1-V^2}}{\sqrt{1-V_z}}\left\{\mbox{\boldmath$d$}(0)+ \left(\,d_z-\frac{1\!-\!\sqrt{1-V^2}}{V^2}\, \left[V_z\,d_z+\mbox{\boldmath$V$}\!_\perp\mbox{\boldmath$\cdot\,d$}(0) \right] \right) \frac{\mbox{\boldmath$V$}\!_\perp}{1-V_z}\right\}\;,
\label{eq:separation}
\eeq
where $d_z\equiv R_z/[D_L\theta_E(0)]$ is the $z$-component of the binary-lens separation vector (the ``depth" of the binary lens along the line of sight) re-scaled by the Einstein radius of the lens. In the following section we study the velocity effect on these parameters, in order to assess its influence on the inferred physical parameters of the microlenses.

\section{VELOCITY EFFECTS ON INFERRED LENSING PARAMETERS}

Light-curve analysis of simple microlensing events does not directly yield the angular scale of the event geometry. Event parameters such as the Einstein radius crossing-time and the impact parameter are obtained scaled to $\theta_E$, and thus will be affected in the same way as the Einstein radius. In addition, the scaled lens separation in the binary-lens case will be affected as shown by equation~(\ref{eq:separation}).

In events beyond the simple model, such as caustic-crossing (source-transit) events or parallax events as well as in events observed astrometrically, it is also possible to measure the angular scale. Therefore, in such events the effect on the angular Einstein radius and the angular lens separation given by expression~(\ref{eq:angularsep}) is potentially of interest.

In the following subsections we study the effects of the lens velocity on the angular Einstein radius $\theta_E(\mbox{\boldmath$V$})$, the angular lens separation $\mbox{\boldmath$\theta$}_{AB}(\mbox{\boldmath$V$})$, and the Einstein-radius scaled lens separation $\mbox{\boldmath$d$}(\mbox{\boldmath$V$})$. In \S3.1 we explore the low-velocity case, which is astrophysically relevant for typical Galactic microlensing settings. In \S3.2 we demonstrate the results for the high-velocity regime.

\subsection{Low Velocities}

The second-order $V\ll1$ expansions of expressions~(\ref{eq:Einstein}), (\ref{eq:angularsep}), and (\ref{eq:separation}) are
\begin{eqnarray}
& & \theta_E(\mbox{\boldmath$V$})\simeq\theta_E(0) \left[\,1- \frac{V_z}{2}+\frac{2V_\perp^{2}+V_z^2}{8}\right]\nonumber \\
& & \mbox{\boldmath$\theta$}_{AB}(\mbox{\boldmath$V$})\simeq \mbox{\boldmath$\theta$}_{AB}(0)+\frac{1}{D_L} \left\{R_z\mbox{\boldmath$V$}\!_\perp+ \frac{1}{2}\left[\,V_z\,R_z-\mbox{\boldmath$V$}\!_\perp \mbox{\boldmath$\cdot R$}_\perp\right] \mbox{\boldmath$V$}\!_\perp \right\} \\
& & \mbox{\boldmath$d$}(\mbox{\boldmath$V$})\simeq \mbox{\boldmath$d$}(0)\left[\,1+\frac{V_z}{2}+ \frac{V_z^2-2V_\perp^2}{8}\right]+ \mbox{\boldmath$V$}\!_\perp\left[(1+V_z)\,d_z-\frac{1}{2}\, \mbox{\boldmath$V$}\!_\perp\mbox{\boldmath$\cdot\,d$}(0)\right] \nonumber\;,
\end{eqnarray}
where $\mbox{\boldmath$\theta$}_{AB}(0)=\mbox{\boldmath$R$}_\perp/D_L$. The angular Einstein radius has a linear order effect only if the lens has a non-zero radial velocity, while any purely tangential velocity produces a non-zero second order effect. As expected from previous results, the radius increases for lenses moving away from the observer and decreases for those moving towards the observer. However, to get a $10^{-3}$ effect on $\theta_E$ would require an unlikely radial velocity $c V_z=600\,km s^{-1}$. We conclude that the effect of lens velocity on the angular Einstein radius is not observationally significant.

From equation~(\ref{eq:angularsep}) we can see that any purely radial velocity of the lens has no effect on the \emph{angular} separation of the lenses $\mbox{\boldmath$\theta$}_{AB}$. The angular separation vector has a linear order effect only if the binary lens has a non-zero tangential velocity and at the same time its axis is tilted from the plane of the sky.  The effect is oriented along the vector of the tangential velocity, its direction depends on the sign of $R_z$. The magnitude of the effect $[\theta_{AB}(\mbox{\boldmath$V$})/\theta_{AB}(0)-1]$ is $R_z(\mbox{\boldmath$V$}\!_\perp \mbox{\boldmath$\cdot R$}_\perp)/R^2_\perp$ with a maximum value of $V_\perp\tan\zeta$ for a purely tangential velocity parallel to the projected binary-lens axis. A low value of $V_\perp$ can thus be offset by a high angle of inclination of the binary-lens axis. For a tangential velocity $c V_\perp=200\,km s^{-1}$ we get a 1\% effect for $\zeta\doteq86^\circ$, which corresponds for example to a projected separation $R_\perp\approx1 AU$ for a binary lens with a physical separation $R\approx15 AU$. The geometric conditions required for even higher effects are no less realistic. We see that the effect is observationally significant for binary-star or star+planet lenses aligned nearly along the line of sight.

To investigate the effect of velocity on the size of the Einstein-radius scaled lens separation vector $d(\mbox{\boldmath$V$})$ we have to treat separately binary lenses oriented along the line of sight, for which $d(0)=0$. In this case we get for the first-order absolute effect
\beq
d(\mbox{\boldmath$V$})-d(0)\simeq |\,d_z\mbox{\boldmath$V$}\!_\perp|\;.
\label{eq:abseffect1}
\eeq
For a given total velocity $V$ the maximum effect $V|d_z|$ occurs for a purely tangential velocity. The zero minimum effect occurs for a purely radial lens velocity.

For a general binary lens not oriented along the line of sight $d(0)>0$ and the first-order absolute effect is
\beq
d(\mbox{\boldmath$V$})-d(0)\simeq \frac{V_z}{2}\;d(0)+ \frac{\mbox{\boldmath$V$}\!_\perp\mbox{\boldmath$\cdot\,d$}(0)} {d(0)}\;d_z\;.
\label{eq:abseffect2}
\eeq
A straightforward computation shows that the maximum value of $V\sqrt{d^2(0)+4\,d_z^2}/2$ is achieved for a velocity orientation
\beq
\mbox{\boldmath$V$}=\frac{V}{R_z\sqrt{1+4R_z^2/R_\perp^2}} \left[ \left( 2\,\frac{R_z^2}{R_\perp^2}-1 \right)\!\mbox{\boldmath$R$}_\perp + \mbox{\boldmath$R$}\, \right]\;.
\label{eq:maxvel}
\eeq
The minimum value has the same effect with a negative sign and it occurs for an opposite velocity, while a zero effect occurs for velocities perpendicular to velocity~(\ref{eq:maxvel}). It is interesting to note that the maximum effect on the scaled lens separation occurs generally for a different velocity orientation than the maximum effect on the angular lens separation. Figure~\ref{fig:velocity} illustrates the dependence of the absolute first-order effect on the direction of the lens velocity for different orientations of the binary-lens axis. In terms of lensing parameters the maximum first-order absolute effect for a binary lens with its axis tilted by $\zeta$ from the plane of the sky is   
\beq
d(\mbox{\boldmath$V$})-d(0)\simeq 0.01 \frac{\sqrt{1+3\sin^2{\zeta}}}{2}\left(\frac{c\,V}{200\,km\,s^{-1}}\right) \left(\frac{\theta_E(0)}{1\,mas}\right)^{-1} \left(\frac{R}{60\,AU}\right)\left(\frac{D_L}{4\,kpc}\right)^{-1}
\eeq
for a velocity direction given by equation~(\ref{eq:maxvel}). The effect for higher lens velocities, wider intrinsic lens separations $R$, and lower Einstein radii can thus be a significant fraction of the Einstein radius. The maximum relative effect $[\,d(\mbox{\boldmath$V$})/d(0)-1\,]$ is $V\sqrt{\cos^{-2}\zeta-0.75}$, which is largest for \mbox{$\zeta\rightarrow90^\circ$}. In this regime it coincides with the maximum relative effect for the angular separation $\theta_{AB}$ derived above.

To summarize the low-velocity results, first-order effects are caused by the radial velocity of the lens and/or the ``depth" of the binary lens along the line of sight. In the case of a single lens the effect of lens velocity on the inferred lensing parameters is not significant. In the case of binary-star or star+planet lenses the inferred values of the angular and Einstein-radius-scaled lens separations (as well as the linearly related lens positions) can differ from the values at zero velocity by more than 1\% if the axis of the lens system is oriented close to the line of sight. In particular, the effect on the scaled lens separation can be a significant fraction of the Einstein radius mainly for lenses with wide intrinsic separations.

\subsection{High Velocities}

The general dependence of the angular Einstein radius on the lens velocity as given by equation~(\ref{eq:Einstein}) is illustrated in Figure~\ref{fig:Einstein}. We can see that the high-velocity regime depends on the orientation -- the case when the lens moves directly towards the observer has to be treated separately.

If we increase the velocity $V\rightarrow 1$ in any other direction (including directly away from the observer), the angular Einstein radius eventually diverges $\theta_E(\mbox{\boldmath$V$})\rightarrow\infty$. For the angular lens separation equation~(\ref{eq:angularsep}) gives us a finite result, and the scaled lens separation thus vanishes $d(\mbox{\boldmath$V$})\rightarrow 0$. Hence, such a ``binary lens" would in effect behave like a single lens.

We note that in this regime the approximation of first order in deflection used in this paper eventually breaks down. However, the divergence of the Einstein radius is very slow -- on the order of $\sim(1-V)^{-1/4}$. From Figure~\ref{fig:Einstein} we can see that even for $V=0.9$ the Einstein radius is at most greater by a factor of two (if the lens moves directly away). Even with a factor of 100 or more the approximation remains valid. It is thus sufficient to formally ignore the results for $V=1$.

A lens moving with a velocity $V\rightarrow 1$ directly towards the observer will have a vanishing angular Einstein radius $\theta_E(\mbox{\boldmath$V$})\rightarrow 0$. The angular lens separation $\mbox{\boldmath$\theta$}_{AB}\rightarrow\mbox{\boldmath$R$}_\perp/D_L$ -- the lenses appear to be at their ``true" positions. The scaled lens separation thus diverges $d(\mbox{\boldmath$V$})\rightarrow\infty$. The decreasing Einstein radius means that such a ``binary lens" would behave like two single lenses with decreasing strength, as the lens moves together with the arriving photons.

\section{DISCUSSION}

In \S3.1 we have demonstrated the low-velocity effects for different binary-lens orientations and intrinsic separations. Nevertheless, from the observational perspective we must take into account the fact that not all binary-lens configurations can lead to microlensing events detectable as two-point-mass lens events. While this depends on the exact source trajectory, from a statistical point of view the Einstein-radius scaled lens separation $d$ plays the main role. If the separation is too small or too large, most events would appear as single-lens events.

For the event to be detectable as a two-point-mass lens event the lens separation has to fulfil $d(0)\in(d_{min},d_{max})$. This leads to limits on the binary-lens axis orientation
\beq
\cos\zeta\in(d_{min},d_{max})\times\left(\frac{R}{4\,AU}\right)^{-1} \left(\frac{D_L}{4\,kpc}\right)\left(\frac{\theta_E(0)}{1\,mas}\right)\;.
\label{eq:cosine}
\eeq
The values of the binary separations from the 21 binary events published by the MACHO team \citep{alc00} range from 0.421 to 2.077 with an outlier value of 7.454 (a possible binary-source event). The 18 events detected by OGLE-II \citep{jaro02} have values from 0.355 to 2.917, and the values for the 15 events detected by OGLE-III \citep{jaro04} range from 0.352 to 3.457.

To obtain a rough estimate we set $d_{min}=0.3$ and $d_{max}=4$. Keeping $D_L$ and $\theta_E$ fixed at the values used in equation~(\ref{eq:cosine}), for an intrinsic binary separation $R=15 AU$ we get limits on the axis angle $\zeta\in(0^\circ,85^\circ)$, for $R=60 AU$ we get $\zeta\in(75^\circ,89^\circ)$, and for $R=240 AU$ we get $\zeta\in(86^\circ,89.7^\circ)$. In the first case we have a wide range of possible orientations -- however, if we approach $90^\circ$ where the velocity effects are strongest, it would be difficult to detect such an event as a binary-lens event. The two cases with higher intrinsic separations demonstrate that such lenses can have stronger velocity effects while being detectable as binary events, albeit for a narrower range of axis orientations.

In this work we concentrated on the center-of-mass velocity of the binary lens and neglected its orbital velocity. However, for binaries or star+planet systems with semi-major axis $\ll 1 AU$ the assumption of small orbital velocity breaks down. The results of this paper indicate that the velocity effects are proportional to the intrinsic lens separation, being caused by the lens motion during the passage of the light ray in its vicinity. While the orbital velocity grows as $R^{-1/2}$ with decreasing $R$, its product with the separation decreases as $R^{1/2}$. We conclude that the effects of orbital lens velocity on light deflection are not significant even in binary-lens systems closer than $1 AU$.

\acknowledgements

This work was supported by grant GACR 205/04/P256 from the Czech Science Foundation.

\clearpage
\bef[t]
\plotone{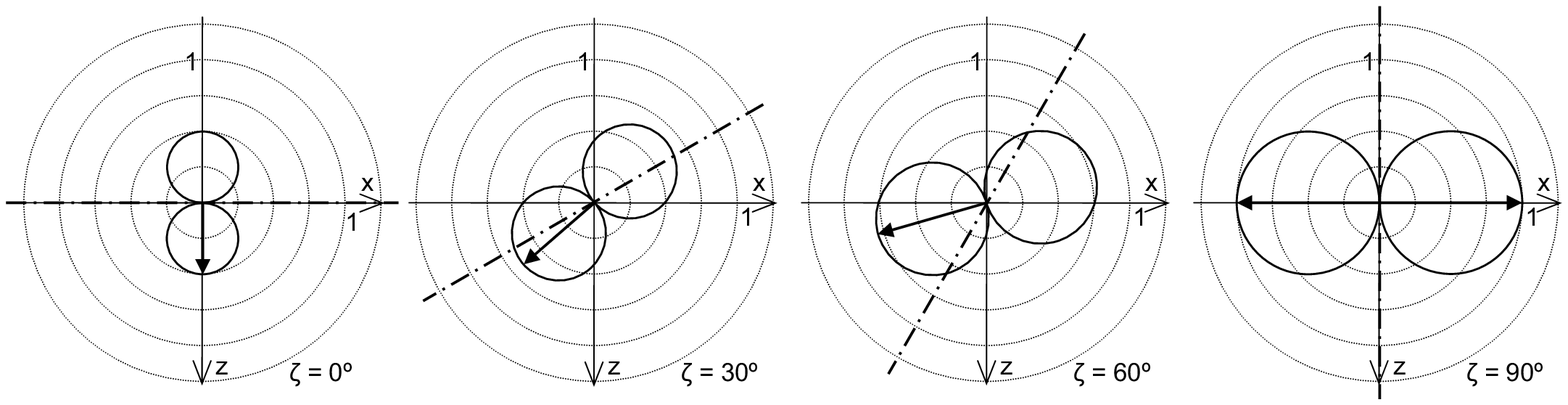}
\caption{Radial plots of velocity effect on binary-lens separation $\,d(\mbox{\boldmath$V$})-d(0)$ given by equations~(\ref{eq:abseffect1}) -- (\ref{eq:abseffect2}), depending on velocity direction in the observer+binary plane ($z$ axis points to observer; $x$ axis in the plane of the sky). Individual plots are for different binary-axis orientations $\zeta$ (dot-dashed lines: binary axis). Effect radially scaled in units of $V\sqrt{d^2(0)+d^2_z}$; arrows mark direction of maximum positive effect given by equation~(\ref{eq:maxvel}). Three-dimensional extensions of the plots are rotationally symmetric: for $\zeta\neq90^\circ$ around the arrow, for $\zeta=90^\circ$ around the $z$ axis.}
\label{fig:velocity}
\enf

\clearpage
\bef[t]
\plotone{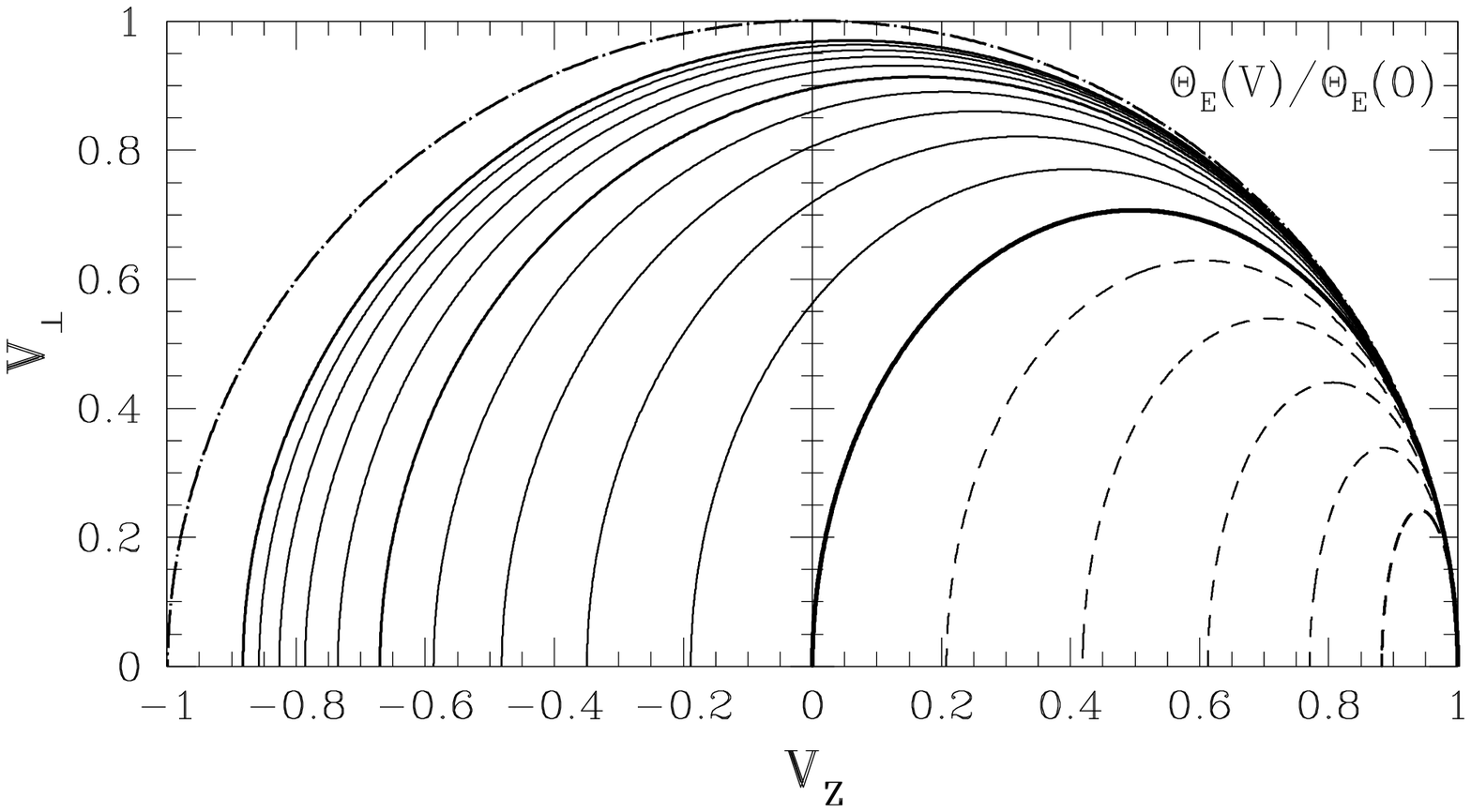}
\caption{Contour plot of angular Einstein radius ratio $\theta_E(\mbox{\boldmath$V$})/\theta_E(0)$ as a function of radial and tangential lens velocities $V_z$ and $V_\perp$ ($V_z$ positive for motion towards observer). Plotted contour values from 0.5 (lower right) to 2 spaced by 0.1: dashed for values $<1$; solid bold for 1; solid for values $>1$. Dot-dashed curve: $V=1$ (speed of light).}
\label{fig:Einstein}
\enf

\end{document}